\documentclass[aps,prb,twocolumn,superscriptaddress,floatfix]{revtex4}
\usepackage{epsfig,amsmath,amssymb,color}
\begin{document}

\title{Thermal transport of the  $XXZ$ chain in a magnetic field}
\author{F. Heidrich-Meisner}
\email{f.heidrich-meisner@tu-bs.de}
\affiliation{Technische Universit\"at Braunschweig, Institut f\"ur Theoretische
  Physik, Mendelssohnstrasse 3, 38106 Braunschweig, Germany}
\author{A. Honecker}
\affiliation{Technische Universit\"at Braunschweig, Institut f\"ur Theoretische
  Physik, Mendelssohnstrasse 3, 38106 Braunschweig, Germany}
\author{W. Brenig}
\affiliation{Technische Universit\"at Braunschweig, Institut f\"ur Theoretische
  Physik, Mendelssohnstrasse 3, 38106 Braunschweig, Germany}

\date{February 15, 2005}

\begin{abstract}
We study the heat conduction of the spin-$1/2$ $XXZ$ chain in finite magnetic fields
where magnetothermal effects arise. Due to the integrability of this model, all transport 
coefficients diverge, signaled by finite Drude weights. Using exact diagonalization 
and mean-field theory, we 
analyze the temperature and field dependence of the thermal Drude weight for
various exchange anisotropies under the condition of zero magnetization-current flow.
First, we find a strong magnetic field dependence of the Drude weight,
	including a suppression of its magnitude with increasing field strength and
a non-monotonic field-dependence of the peak position. Second, 
for small exchange anisotropies and  magnetic fields in the massless 
as well as  in the fully polarized regime
the mean-field approach is in excellent agreement with the exact diagonalization data. 
Third, at the field-induced quantum critical
line between the para- and ferromagnetic region we propose a universal
low-temperature behavior of the thermal Drude weight.
\end{abstract}

\maketitle

\section{Introduction}
Transport properties of one-dimensional spin-$1/2$ systems are currently at the focus
of active  research. This has been motivated by the experimental manifestation 
of   significant contributions to the thermal conductivity originating from magnetic excitations
\cite{solo00,solo01,kudo01,hess01,solo03,hess04}, stimulating intensive theoretical 
work\cite{alvarez02a,hm1,hm2,saito03,orignac03,
shimshoni03,zotos04,hm04,kluemper02,sakai03,rozhkov04,louis03,sakai04}.
Strong theoretical
efforts\cite{alvarez02a,hm1,hm2,saito03,orignac03,zotos04,hm04} 
have been devoted to the question of possible ballistic thermal transport
in generic spin models such as spin ladders, frustrated chains, and dimerized chains. 
Such ballistic transport would be characterized by a finite thermal Drude weight.
Recent numerical and analytical studies
indicate  that in pure but {\it nonintegrable}  spin models, the thermal  Drude
weight scales to zero  
in the thermodynamic limit 
implying that the thermal 
current is likely to have a finite intrinsic life-time\cite{hm1,hm2,shimshoni03,zotos04,hm04}. 
 In addition,
the effects of extrinsic magnon scattering by phonons and/or impurities
have been addressed in several works\cite{orignac03,shimshoni03,rozhkov04}.
For the integrable $XXZ$ model, the energy current operator is a conserved quantity\cite{niemeyer71,zotos97},
leading to a finite thermal Drude weight. Its temperature dependence has been studied with exact diagonalization\cite{alvarez02a,hm1,hm2} and 
 Bethe ansatz techniques\cite{kluemper02,sakai03} and is well understood for arbitrary values of the exchange anisotropy 
at zero magnetic field.
In this paper, we address the issue of thermal transport in the $XXZ$ model  in the presence of 
a finite magnetic field $h$. In this case,  magnetothermal 
effects become important and must be accounted for. 
The magnetothermal response itself has  been
studied  by Louis and Gros in the limit of small magnetic fields\cite{louis03} and recently also
by Sakai and Kl\"umper in the low-temperature limit\cite{sakai04}.
Here, we consider magnetic fields of arbitrary strength and we
discuss the temperature dependence of the thermal 
Drude weight  under the condition of zero spin-current flow. \\
\indent
The Hamiltonian of the $XXZ$ model reads
\begin{equation} H =J \sum_{l=1}^{N} \left\lbrace \frac{1}{2} (S_l^+S_{l+1}^-+\mbox{H.c.})+{\Delta }S_l^{z}S_{l+1}^{z} -h
S_l^z \right\rbrace\label{eq:m2}
     \end{equation}
        where $N$ is the number of sites, $S_{l}^{z,\pm}$ are spin-$1/2$
 operators acting on site $l$, and $\Delta$ denotes  the exchange anisotropy. 
 The exchange coupling $J$ is set to unity in our numerical calculations.
  We focus on $\Delta \geq 0$ and periodic boundary conditions are imposed.\\
  \indent
 \begin{table}[b]
\begin{tabular}{|l|c|c|c|c|}
\hline
 	& 		&$h$					& $m_0$		& $T/J\ll 1$		\\\hline
(i)	& FM, gap	& $h>h_{\mathrm{c2}}$			&	1/2	& $K_{\mathrm{th}}\propto
T^{3/2}\,\mbox{exp}(-G/T)$	\\\hline
	& Saturation	& $h=h_{\mathrm{c2}}$			&	1/2	&$ K_{\mathrm{th}} = \mbox{const} \,T^{3/2}$		\\\hline	
	&		& $h_{\mathrm{c2}}=1+\Delta$		&		&							\\\hline	
(ii)	& Massless	& $h_{\mathrm{c1}}<h<h_{\mathrm{c2}}$	&		& $K_{\mathrm{th}}\propto T$				\\\hline	
(iii)	& AFM, gap	& $ h<h_{\mathrm{c1}}$		&	0	& 	\\\hline	
	\end{tabular}
\caption{
Magnetic phases  of the $XXZ$ model (see, e.g., Ref.~\onlinecite{cabra98}) 
and leading term of the thermal Drude weight $K_{\mathrm{th}}$ at low temperatures.
 $ m_0 $ is the average local magnetization at $T=0$. $G=G(h)$ denotes the gap in either the polarized state (i) or the massive
 antiferromagnetic regime (iii). In the polarized state (i), $G(h)/J= h-h_{c2}$. 
}\label{tab:1}
\end{table}
The quantum phases and the spectrum of (\ref{eq:m2}) are well understood,
both as a function of exchange anisotropy $\Delta$ and magnetic field $h$.
The reader is referred to Ref.~\onlinecite{cabra98} for a detailed summary and further references.
Here we only repeat  the main points.  At zero
magnetic field,  the spectrum of the Hamiltonian Eq.~(\ref{eq:m2}) is gapless
for $|\Delta|\leq 1$ and gapped for $|\Delta| >1$. The situation at finite
magnetic fields is summarized in the first four columns of Table~\ref{tab:1}.
Three different cases are found:
   (i) the ferromagnetic gapped state for $h>h_{c2}=1+\Delta$ (FM); (ii) the gapless or 
   massless phase for $h < h_{c2}=1+\Delta$ and $h>h_{c1}$; 
   and (iii) the antiferromagnetic, gapped
   state for $\Delta>1$ and $h<h_{c1}$ (AFM).
 The line $h=h_{c1}$ starts at the $SU(2)$ symmetric point $\Delta=1, h=0$
   and $h_{c1}$ grows exponentially slowly in the region $\Delta>1$, $h>0$.
  \\\indent
 The fifth column of Table~\ref{tab:1} is a first account of our main findings  
 for the low-temperature behavior of the thermal Drude weight,
   denoted by $K_{\mathrm{th}}$ in this paper. These results are now briefly
   summarized.
 One can expect qualitative changes in 
  the low-temperature behavior of the thermal Drude weight
  as the transition lines $h=h_{c1}$ and $h=h_{c2}$
 are crossed. In particular, we focus on the transition from the gapless phase to the ferromagnetic state.
 In Sec.~\ref{sec:3}, we will argue that for $T/J\ll 1$, first, $K_{\mathrm{th}}\propto T^{3/2}
 \,\mbox{exp}(-G/T)$ 
 in the ferromagnetic state, $G$ being the gap,
 and $T$ temperature;
 second, $K_{\mathrm{th}}\propto T$ in the massless phase; and third, $K_{\mathrm{th}}\propto T^{3/2}$ 
 along the line
 $h=h_{c2}$.
 \\\indent
 Regarding  the antiferromagnetic state,
 there is  certainly also an exponentially suppressed Drude weight; 
 see for instance Refs.~\onlinecite{hm1} and \onlinecite{sakai03} for $h=0$. 
 However, the low-temperature region in this case 
 and for $h=h_{c1}$ is difficult to
 reach with the methods of the present paper. For a discussion of 
 the low-temperature limit at vanishing magnetic field, we refer the reader 
 to Refs.~\onlinecite{hm2,hm1,kluemper02} and \onlinecite{sakai03}. 
 Apart from the low-temperature behavior, this paper studies the field  dependence of the thermal Drude
 weight in the phases (i) and (ii) at finite temperatures.
 

 The plan of this paper is the following. First, we discuss the expressions for the transport coefficients 
 and the current operators in Sec.~\ref{sec:2}. Second, in Sec.~\ref{sec:3} we perform an
 analysis of the transport coefficients based on a Jordan-Wigner
 mapping of the spin system onto spinless fermions. In this case, interactions at
 $\Delta \not= 0$ will be treated by a Hartree-Fock approximation.
 Third, we present our results from exact diagonalization for $\Delta> 0$ in Sec.~\ref{sec:4} and compare 
 them to the results from the Jordan-Wigner approach. 
 The field and temperature dependence of 
 the thermal Drude weight is discussed with a particular focus on the case of the Heisenberg chain.
 A summary and conclusions are given in Sec.~\ref{sec:5}. 


\section{Transport coefficients}\label{sec:2}
Within linear response theory,
the thermal and the spin current are related to the gradients $\nabla h$ 
and $\nabla T$ of the field 
$h$ and the temperature $T$ by\cite{mahan}
\begin{equation}
\left(\begin{array}{c}
J_{1} \\
J_{2}
\end{array}\right) =
\left(\begin{array}{cc}
L_{11} & L_{12} \\
L_{21} & L_{22}
\end{array}\right) 
\left(\begin{array}{c}
\nabla h\\ 
-\nabla T
\end{array}\right)
\label{eq:k0}
\end{equation}
where $J_{i}=\langle j_{i}\rangle $ is either the thermodynamic expectation value
of the spin current $j_{1}$ or the thermal  current operator $j_{2}$, respectively.
$L_{ij}$ denote the transport coefficients. At finite  frequencies $\omega$,
the coefficients $L_{ij}(\omega)$ depend  on the time-dependent current-current correlation
functions via \cite{mahan}
\begin{equation}
L_{ij}(\omega)  
= \frac{\beta^{r}}{N}
 \int_0^{\infty} dt\, e^{i(-\omega+i0^+ )t}
         \int_0^{\beta} d\tau \langle j_{i}\,j_{j}(t+i\tau)\rangle \, .
\label{eq:k100}
\end{equation}
 In this equation, $r=0$ for $j=1$ and $r=1$ for $j=2$. $\beta=1/T$ is the 
inverse temperature and $\langle .\rangle$ denotes the thermodynamic expectation value. 
Note that  $L_{12}= L_{21}/T$ due to Onsager's relation\cite{mahan}.  
The real part of $L_{ij}(\omega)$  
can be decomposed into a  $\delta$-function at $\omega=0$ with weight $D_{ij}$ and a 
regular part $L_{ij}^{\mathrm{reg}}(\omega)$:
\begin{equation}
\mbox{Re}\, L_{ij}(\omega)=
D_{ij} \delta(\omega)+ L_{ij}^{\mathrm{reg}}(\omega)\,.
\label{eq:exp:k101}
\end{equation}
This equation defines the Drude weights $D_{ij}$, for which a spectral representation
can be given\cite{zotos97}
\begin{equation}
D_{ij}(h,T) =
   \frac{\pi\beta^{r+1} }{N} \sum_{m, n\atop E_m=E_n}p_n
     \langle n | j_{i}|m\rangle   \langle m | j_{j}|n\rangle \,. 
\label{eq:k102}
\end{equation}
Here, $p_n=\mbox{exp}(-\beta E_n)/Z$ is the Boltzmann weight and $Z$ denotes the partition function.
In the exponent, $r$ has to be chosen in the same way as in Eq.~(\ref{eq:k100}). 
 \\
\indent
Let us now introduce the appropriate definitions of the current operators.
The local current operators $j_{1,l}$ and $j_{2,l}$
satisfy the continuity equations 
\begin{equation}
j_{j,l+1}  - j_{j,l}	=  -i\lbrack H, d_{j,l}\rbrack ;\quad j=1,2
\label{eq:2b}
\end{equation}
where $d_{1,l}=S_l^z$ is the local magnetization density and $d_{2,l}=h_l$ is the local energy density,
respectively, with $H=\sum_l h_l$. 
At zero magnetic field, the total currents $j_{\mathrm{th[s]}}=\sum_l j_{\mathrm{th[s]},l}$ are given by\cite{zotos97,kawasaki63,shastry90}
\begin{eqnarray}
j_{\mathrm{s}}&=& i J\sum_{l=1}^{N} (S_{l}^+S_{l+1}^- - S_{l+1}^+S_{l}^-)\label{eq:3}\\
j_{\mathrm{th}}&=&  J^2\sum_{l=1}^{N} \tilde{\vec{S}}_{l} \cdot ( \vec{S}_{l+1}\times \tilde{\vec{S}}_{l+2})  \label{eq:4}
\end{eqnarray}
with the definition $\tilde{\vec{S}}_l=(S_l^x,S_l^y,\Delta S_l^z)$to achieve a compact representation, while
$\vec{S}_l$ is defined as usual. Note that subscripts in brackets $\lbrack . \rbrack$ refer to
 spin transport.\\\indent
At finite magnetic field, the proper set of current operators is\cite{louis03} 
\begin{equation}
j_{1}=j_{\mathrm{s}}; \quad j_{2}=j_{\mathrm{th}}-h j_{\mathrm{s}}.
\label{eq:k1}
\end{equation}
\indent
Now, the crucial point is that, while the {\em spin} current $j_{\mathrm{s}}$ is
only conserved  in the $XX$ case 
($\Delta=0$), the  current $j_{\mathrm{th}}$ is conserved  for  all fields
	$h$ and values of $\Delta$, i.e.,  
$\lbrack H,j_{\mathrm{th}}\rbrack = 0 $
(Refs.~\onlinecite{niemeyer71} and \onlinecite{zotos97}).
Thus, it immediately follows from Eqs.~(\ref{eq:k100}) and (\ref{eq:k1})
that the Drude weights $D_{12}$, $D_{21}$, and $D_{22}$ are finite 
for arbitrary fields $h$.\\\indent
Furthermore, one can show that the spin Drude weight $D_{11}$ is also finite 
in the thermodynamic limit for $h\not= 0$.   We briefly outline the proof along the lines of 
Ref.~\onlinecite{zotos97}.
 Given    a 
set of all conserved observables  $\lbrace Q_l\rbrace$,  the spin Drude weight $D_{11}$ can be written as
 \begin{equation}
 D_{\mathrm{11}}(h,T)= \frac{\pi}{T\,N}  ( j_{\mathrm{1}} \,|\, {\cal{P}} j_{\mathrm{1}} )
\label{eq:20}
\end{equation}
where $\cal{P}$ is the projection operator in the Liouville space 
on the subspace spanned by all   conserved quantities $\lbrace Q_l\rbrace$. 
The brackets $(.|.)$ denote
Mori's scalar product; see, e.g., Ref.~\onlinecite{forster} for details. 
Restricting to a subset $\lbrace Q_m\rbrace \subset \lbrace Q_l\rbrace$, one obtains an 
inequality\cite{zotos97,mazur69}
 \begin{equation}
 D_{\mathrm{11}}(h,T)\geq \frac{\pi}{T\,N}  \sum_m \frac{\langle j_{\mathrm{1}}Q_m\rangle^2}{\langle  Q_m^2\rangle},
\label{eq:21}
\end{equation}
providing a lower bound for the Drude weight $D_{11}(h,T)$.  
In the literature, this relation is often referred to as 
Mazur's inequality\cite{zotos97,mazur69}. 
Several authors\cite{zotos97,fujimoto03} have used Eq.~(\ref{eq:21})
to infer a finite spin Drude weight for the Heisenberg chain, assuming
broken particle-hole symmetry, or the presence of a finite magnetic field, respectively.
More explicitly, only one   
conserved quantity is often considered in  Eq.~(\ref{eq:21}),
namely $Q_1=j_{\mathrm{th}}$, which has a finite overlap $(j_{\mathrm{1}}|j_{\mathrm{th}})>0$ with the spin current 
for $h\not= 0$. This finally proves $D_{11}(h,T)>0$ for $h\not= 0$. \\ 
\indent
The main focus of this paper is on  the case of  purely thermal transport with a vanishing spin
	current, i.e.,  $J_{1}=0$. We therefore arrive at a 
thermal conductivity $\kappa$ which is described by
\begin{equation}
\mbox{Re}\,\kappa(\omega)=K_{\mathrm{th}}(h,T)\delta (\omega)+
\kappa_{\mathrm{reg}}(\omega)\,
\end{equation}
where $K_{\mathrm{th}}(h,T)$ in terms of the Drude weights $D_{ij}$ reads
 \begin{equation}
K_{\mathrm{th}}(h,T) = D_{22}(h,T) - \beta \frac{D_{21}^2(h,T)}{D_{11}(h,T)}\,.
\label{eq:k5}
\end{equation}
 Exactly the same result  for $K_{\mathrm{th}}(h,T)$ is obtained if a different choice of current operators 
and corresponding forces is made, e.g., $j_{\mathrm{s}}$ and $j_{\mathrm{th}}$ from Eqs.~(\ref{eq:3}) and (\ref{eq:4}) 
(see Ref.~\onlinecite{mahan}). 
The expression for $K_{\mathrm{th}}(h,T)$, being fully equivalent to Eq.~(\ref{eq:k5}), is then given by
 \begin{equation}
K_{\mathrm{th}}(h,T) = D_{\mathrm{th}}(h,T) - \beta \frac{D_{\mathrm{th,s}}^2(h,T)}{D_{\mathrm{s}}(h,T)}.
\label{eq:k5a}
\end{equation} 
Note that for $h=0$, $K_{\mathrm{th}}(h=0,T)=D_{\mathrm{th}}(h=0,T)$.
Therefore, two competing terms contribute to $K_{\mathrm{th}}(h,T)$ in Eq.~(\ref{eq:k5a}): the "pure" thermal Drude weight $D_{\mathrm{th}}$
and the "magnetothermal correction",  $\beta D_{\mathrm{th,s}}^2/D_{\mathrm{s}}$. Note that 
the magnetothermal correction might be suppressed by external scattering or spin-orbit coupling, breaking the conservation 
of the total magnetization of the spin system (Ref.~\onlinecite{shimshoni03}). This is an open issue which 
may depend crucially on the particular material investigated  in experimental transport studies.
\\
\indent
Let us now give spectral representations for the  quantities $D_{\mathrm{s}}(h,T)$, 
$D_{\mathrm{th}}(h,T)$, and $D_{\mathrm{th,s}}(h,T)$ (Refs.~\onlinecite{kohn64,castella95,zotos97}) 
\begin{eqnarray}
D_{\mathrm{th[s]}}^I(h,T) &=&
   \frac{\pi \beta^{2[1]}}{N} \sum_{m, n\atop E_m=E_n} p_n
      |\langle m| j_{\mathrm{th[s]}}|n\rangle|^2,\label{eq:k3}\\
D_{\mathrm{s}}^{II}(h,T) &=& \hspace{-0.1cm}\frac{\pi}{N}\left\lbrack \langle  - \hat T\rangle -2
\hspace{-0.3cm}\sum_{m, n\atop E_m\not=E_n}\hspace{-0.2cm}
 p_n\frac{|\langle m |
j_{\mathrm{s}}|n\rangle|^2}{E_m-E_n} \right\rbrack\hspace{-0.1cm},\label{eq:k8}\\
D_{\mathrm{th,s}}(h,T)&=& \frac{\pi\beta}{N} \sum_n  p_n \langle n| j_{\mathrm{th}} j_{\mathrm{s}}|n\rangle.
\label{eq:k4}
\end{eqnarray}
The operator $\hat T =(1/2)\sum_l
(S_l^+S_{l+1}^-+\mbox{H.c.})$ is the kinetic energy.
 In the Eqs.~(\ref{eq:k3}), (\ref{eq:k8}), and (\ref{eq:k4}), the magnetic field only enters via 
the Boltzmann weights $p_n$.
The two expressions $D_{\mathrm{s}}^{I}$ and $D_{\mathrm{s}}^{II}$ are
equivalent in the thermodynamic limit, but exhibit differences
at low temperatures for finite system sizes\cite{scalapino93,giamarchi95,zotos97,narozhny98,kirchner99}. 
In this context, note that $D_{\mathrm{s}}^{II}-D_{\mathrm{s}}^{I}$ is the so-called Meissner fraction, 
which measures the {\it superfluid}
density in the thermodynamic limit and in a transverse vector-field\cite{scalapino93,giamarchi95}. This quantity vanishes for $N\to\infty$ in one dimension, but it can be nonzero
for finite systems\cite{giamarchi95}. In Ref.~\onlinecite{hm2}, we have performed a  study of the finite-size scaling
of both quantities for the $XXZ$ chain, showing that $D_{\mathrm{s}}^{I}\approx D_{\mathrm{s}}^{II}$
already holds at sufficiently high temperatures. 
At low temperatures and zero magnetic field, $D^I_{\mathrm{s}}$ is always exponentially suppressed  for even $N$ due to 
finite-size
gaps; thus a finite value  of   $D_{\mathrm{s}}(T=0)$ can only be found
for $N\to \infty$. On the contrary, since $D_{\mathrm{s}}^{II}\approx (\pi/N)\langle -\hat T\rangle$ at low
temperatures,
$D_{\mathrm{s}}^{II}$ correctly results in a {\it finite} value at $T=0$ in the massless regime.
Depending on the context, one should carefully check
which of these two quantities exhibits the more reliable finite-size behavior, and 
in fact, in the present case of finite magnetic fields we will argue in Sec.~\ref{sec:4}
that  $D_{\mathrm{s}}^I$ should preferably  be used.
For a more detailed discussion of the relation between $D_{\mathrm{s}}^{I}$ and $D_{\mathrm{s}}^{II}$, we refer the reader to
 Ref.~\onlinecite{hm2} and references therein.\\
 \indent
In our numerical analysis, we will evaluate $D_{\mathrm{th}}$, $D_{\mathrm{s}}$, and $D_{\mathrm{th,s}}$
while the coefficients $D_{ij}$ from Eq.~(\ref{eq:k5}) 
can   be derived if desired  as they are linear combinations of 
 $D_{\mathrm{th}},D_{\mathrm{s}}$, and
$D_{\mathrm{th,s}}$:
\begin{eqnarray}
D_{11}&=& D_{\mathrm{s}},  \label{eq:k9a}\\
D_{21}&=& D_{\mathrm{th,s}}-h D_{\mathrm{s}},\label{eq:k9b}\\
\quad D_{22}&=& D_{\mathrm{th}} -2\,\beta h D_{\mathrm{th,s}}+\beta h^2 D_{\mathrm{s}}.
 \label{eq:k9c}
\end{eqnarray}

\indent
The $XXZ$ model is integrable and solvable via the Bethe
	ansatz. Therefore one expects all quantities in Eqs.~(\ref{eq:k5}) and
	(\ref{eq:k5a}) to be accessible by analytical techniques.
Yet, for the spin Drude weight $D_{\mathrm{s}}(h=0,T)$ at zero magnetic field, partly
contradicting results can be found in the literature regarding both its
temperature dependence and the question
whether it is finite or not for the Heisenberg chain ($\Delta=1$)  in the thermodynamic limit.
See Refs.~\onlinecite{hm2,castella95,zotos96,narozhny98,zotos99,alvarez02b,long03,fujimoto03} 
and further references therein.


\section{Mean-field approximation}
\label{sec:3}

We now discuss a Hartree-Fock type of approximation to the Hamiltonian
Eq.~(\ref{eq:m2}), which we use to compute the Drude weights $D_{ij}$. 
The spin operators $S^{z,\pm}_l$ are first mapped 
onto spinless fermions via the Jordan-Wigner transformation\cite{mahan}
\begin{equation}
S^z_l=c_l^{\dagger}c_l^{ }-\frac{1}{2}; \quad S^+_l=e^{i\pi\Phi_l}c_l^{\dagger}.
\label{eq:m4}
\end{equation}
Here, $c_l^{(\dagger)}$ destroys(creates) a fermion on site $l$. The string operator $\Phi_l$ reads
$
\Phi_l=\sum_{i=1}^{l-1}\, n_i
$ with $n_i=c_i^{\dagger}c_i^{ }$.
Next, the interaction term $\Delta n_{l}n_{l+1}$ appearing in the fermionic representation 
is treated by a Hartree-Fock decomposition leading to an effective
mean-field Hamiltonian
\begin{equation}
H_{\mathrm{MF}}=\sum_k \epsilon_k c^{\dagger}_k c^{ }_k
\label{eq:mf1}
\end{equation}
with the mean-field dispersion 
\begin{equation}
\epsilon_k = -J\lbrace (1+2 \Delta \alpha)\cos(k) +h- 2 \Delta (n-1/2)\rbrace \, .
\label{eq:mf2}
\end{equation}
The quantities to be determined self-consistently are 
 $\alpha=\langle c_{l+1}^{\dagger} c^{ }_l \rangle$ and  $n=\langle c_{l}^{\dagger} c^{ }_l \rangle$
 where the latter is related to the average local magnetization $m$ via 
 $m=\langle S_{l}^z \rangle= n-1/2$.
The Drude weights can then be obtained from 
\begin{eqnarray}
D_{11} &=&(\pi \beta/N) \langle j_1^2\rangle; \label{eq:mf3}\\ 
D_{21} &=&(\pi\beta  /N) \langle j_2j_1\rangle;\label{eq:mf4}\\ 
 D_{22}&=&(\pi \beta^2/N) \langle j_2^2\rangle \label{eq:mf5}.
\end{eqnarray}
The current operators read
$$
j_1=\sum_k v_k c_k^{\dagger} c_k^{};\quad j_2=\sum_k \epsilon_k v_k  c_k^{\dagger}c_k^{}
$$
with $v_k=d\epsilon_k/dk$.\\\indent
While this approach is  exact for $\Delta=0$, fair results for
$K_{\mathrm{th}}(h=0,T)$ are even obtained
for $0<\Delta\leq 1$; see Refs.~\onlinecite{hm1} and \onlinecite{hm2}. 
From Eqs.~(\ref{eq:mf3}) to (\ref{eq:mf5}), the leading contribution at low temperatures  
can be derived. \\\indent
We start with the free  fermion case $\Delta=0$, for which
we find at the saturation field $h_{c2}$
\begin{equation}
K_{\mathrm{th}}(h,T)=\left( A_{\mathrm{22}} -\frac{A_{\mathrm{21}}^2}{A_{\mathrm{11}}}\right) T^{3/2}\enspace
\mbox{for}\enspace h=h_{c2}
\label{eq:mf7}
\end{equation}
with
\begin{eqnarray}
A_{11}&=& \sqrt{\frac{{\pi}}{2}}(1-\sqrt{2})\, \zeta(1/2)\nonumber, \\
A_{21}&=& \frac{3}{4}\sqrt{\frac{{\pi}}{2}}(2-\sqrt{2}) \,\zeta(3/2),\label{eq:mf9}\\
A_{22}&=& \frac{15}{16}\sqrt{\frac{{\pi}}{2}} (4-\sqrt{2})\, \zeta(5/2); \nonumber 
\end{eqnarray}
$\zeta(x)$ being the Riemann-Zeta function. 
Note that the spin Drude weight at $T=0$ is finite for $0<h<h_{c2}$ and vanishes
for $h\geq h_{c2}$. At low temperatures and for $h=h_{c2}$, 
we find $D_{\mathrm{11}}(T)= A_{11}\,\sqrt{T}$ and a  
divergence of the pure thermal Drude weight $D_{\mathrm{th}}$ with  $D_{\mathrm{th}}\approx h_{c2}^2 A_{11} T^{-1/2} $ to leading
order in temperature, which follows from 
Eqs.~(\ref{eq:k9b}) and (\ref{eq:k9c}).
 We mention that 
 the result $D_{22}\propto T^{3/2}$  at the critical field was also  found 
  within a
 continuum theory  suggested to describe transport properties of two-leg spin ladders.\cite{orignac03}  \\
\indent
In the intermediate regime, i.e., the gapless state (ii) [see Table~\ref{tab:1}],
\begin{equation} K_{\mathrm{th}}(h,T)=\frac{\pi^2}{3} v(h) \, T; \quad v(h)=J\,\sqrt{1-h^2}
\label{eq:mf9a}
\end{equation}
holds at low temperatures,  because the dispersion is  linear
in the vicinity of the Fermi level for $k_{\mathrm{F}}\not= 0,\pi$. 
Note that $K_{\mathrm{th}}(h,T) \approx D_{\mathrm{22}}(h,T)$
for small $T$ in this regime. Equation~(\ref{eq:mf9a}) results 
in 
$
K_{\mathrm{th}}={\pi^2 J T}/{3} 
$ for $h=0$, which is, e.g., known from Ref.~\onlinecite{kluemper02}. 
\\ 
\indent
 For $|h|> |h_{c2}|=1$, both 
$D_{\mathrm{22}}$ and the second term in Eq.~(\ref{eq:k5}), i.e., $D_{21}^2/(T\,D_{11})$,
are given by
\begin{equation}
D_{22}= D_{21}^2/(T\,D_{11})=\sqrt{\frac{\pi}{2}}\, G^2\, \frac{e^{-G/T}}{\sqrt{T}}\,,
\end{equation}
to leading order in temperature and for $T \ll G$, where $G/J=|h|-1$ is the gap. 
This implies that $K_{\mathrm{th}}(h,T)$
is strongly suppressed at low temperatures due to the cancellation of the contributions
to $K_{\mathrm{th}}(h,T)$ in Eq.~(\ref{eq:k5}). In fact, 
such cancellation occurs in the next-to-leading order in $T$ as well. 
One can further show, taking into account the first non-vanishing contribution to $K_{\mathrm{th}}$ 
in Eq.~(\ref{eq:k5}), that 
\begin{equation}
K_{\mathrm{th}}(h,T)= \frac{3}{4}\,\sqrt{{2\,\pi}}\,  T^{3/2} \,e^{-G/T}
\end{equation}
describes the low-temperature behavior of the thermal Drude weight above $h_{c2}$.
In Ref.~\onlinecite{orignac03}, it has  been argued that  $D_{22}\propto \mbox{exp}(-G/T)/\sqrt{T}$
 is a generic feature of  gapped systems with a finite thermal Drude
 weight.\\\indent
 We further point out that   
 the ratio of the thermal Drude weight $K_{\mathrm{th}}$ and the spin Drude weight $D_{\mathrm{s}}$ fulfills a Wiedemann-Franz type of
 relation in the low-temperature limit in all three cases, i.e., in the massless and the fully polarized state as well as for $h=h_{c2}$:
 \begin{equation}
 \frac{K_{\mathrm{th}}}{D_{s}}= L_0\, T\,.
 \label{eq:100}
 \end{equation}
 The constant $L_0$ takes  different values in the regimes (i) and (ii), but for the free-fermion case 
 (and within mean-field theory as well) it is independent of the magnetic field
 {\it in} the massless and fully polarized state, respectively.
  \\
\begin{figure}[t]
\centerline{\epsfig{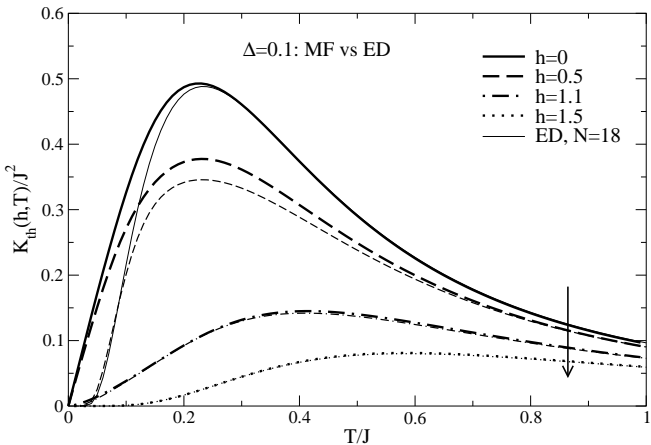}}
\caption{
Thermal Drude weight $K_{\mathrm{th}}(h,T)$ of the $XXZ$ chain for $\Delta=0.1$:
comparison of mean-field theory (MF) and exact diagonalization (ED).
The thermal Drude weight $K_{\mathrm{th}}(h,T)$ is shown for
$h=0,0.5,1.1,1.5$.
Increasing field is indicated by the arrow. Thick  lines denote 
results from the Hartree-Fock approximation; thin lines: ED for $N=18$. 
Deviations at low temperatures for $h=0$ and $h=0.5$ are due to finite-size effects in the ED
results.
}\label{fig:1}
\end{figure}
\indent
 Before turning to the mean-field theory for $\Delta>0$,  let us briefly discuss
 which results can be expected from  conformal field theory for the massless state.
 The expressions for the spin and thermal Drude weight
$D_{\mathrm{22}}$ and $D_{\mathrm{11}}$
have the same structure as at zero magnetic field, i.e.,
$
D_{\mathrm{22}}(h,T)=(\pi^2/3) v(h,\Delta)\, T  
$
and $D_{\mathrm{11}}(h,T)=  K(h,\Delta)\, v(h,\Delta) $ with 
field-dependent velocity $v$ and Luttinger parameter $K$ (see, e.g., Refs.~\onlinecite{hm1} and \onlinecite{hm2}). 
This implies that the constant $L_0$ appearing in Eq.~(\ref{eq:100}) is field dependent 
in the massless regime (see Ref.~\onlinecite{kluemper02} for $h=0$):
\begin{equation}
L_0= \frac{\pi^2}{3\,K(h,\Delta)}\,.
\end{equation}
Furthermore, $D_{21}$ vanishes in the continuum limit due to particle-hole symmetry. While a finite magnetic
field initially breaks this symmetry for the original bosonic fields, the original form of the Luttinger-liquid
Hamiltonian is restored by introducing a shifted bosonic field\cite{cabra98}.
This has an interesting consequence for the low temperature behavior of the pure thermal 
Drude weight $D_{\mathrm{th}}$. Namely,  by solving Eqs.~(\ref{eq:k9b})
and ({\ref{eq:k9c}}) for $D_{\mathrm{th}}$, one obtains
\begin{equation}
D_{\mathrm{th}}=  D_{22}+ \frac{h^2 D_{\mathrm{s}}}{T}=\frac{\pi^2}{3} v T + K v \frac{h^2}{T^{-1}} \,,
\label{eq:k99}
\end{equation}
which implies that $D_{\mathrm{th}}$ diverges at low temperatures with $T^{-1}$ in the massless regime,
consistent with results of Ref.~\onlinecite{sakai04}.\\\indent
Additionally, one obtains $K_{\mathrm{th}}$ in the massless regime and in the low-temperature limit
\begin{equation}
K_{\mathrm{th}}(h,T)\approx D_{22}=\frac{\pi^2}{3} v(h,\Delta)\, T \,.
\label{eq:xx}
\end{equation}
Both parameters, i.e., $K=K(h,\Delta)$ and $v=v(h,\Delta)$, can be computed exactly by solving the Bethe-ansatz 
equations\cite{bogoliubov86}. 
The velocity $v=v(h)$ has been calculated for $\Delta=1$ in Ref.~\onlinecite{hammar99}. 
Further numerical values 
for  these parameters can be found in, e.g., Ref.~\onlinecite{qin97}.
\\\indent 
Let us next dicuss the results from the mean-field approximation (MF)
for $\Delta>0$. 
Figure~\ref{fig:1} shows $K_{\mathrm{th}}(h,T)$ for $\Delta=0.1$ and $h=0,0.5,1.1,1.5$ (thick lines).
The main features are: (i) a suppression of the thermal Drude weight by the magnetic field; (ii) 
a shift of the maximum to higher temperatures for  $h > 0.5$ compared to $h=0$; (iii) a change in the 
low-temperature behavior which will be discussed in more detail below in this section.\\
\indent
For comparison, the results from exact diagonalization (ED) for $N=18$ sites are included  
in Fig.~\ref{fig:1} 
(thin lines) 
and we find that the agreement is very good. Deviations at low temperatures for $h =0$ and $h=0.5$ are due to 
finite-size effects, i.e., the ED results are not yet converged to the thermodynamic limit. For larger fields  
$h \geq  h_{c2}=1.1$, deviations between ED and MF are negligibly small.\\
\indent
From Eq.~(\ref{eq:mf2}), we can derive the critical field $h_{c2}$ within the Hartree-Fock approximation.
At $T=0$ and $h=h_{c2}$, the ground state is the fully polarized state 
with $n=\langle c^{\dagger}_i c_i^{ }\rangle=1$, 
 i.e.,  the  parameter $\alpha$ from Eq.~(\ref{eq:mf2})
 vanishes. Consequently, we find $h_{c2}=1+\Delta$ in accordance with the 
exact result\cite{cabra98}. 
Indeed, the low-energy theories along the line $h=1+\Delta$ and for $\Delta=0$ are equivalent in the sense  that 
they are 
characterized 
by the same  Luttinger parameter\cite{bogoliubov86,dzhaparidze78}.
Within bosonization, the 
line $h=h_{c2}$ is particular since the velocity of the elementary excitations vanishes here.  \\
\indent
Regarding the low-temperature behavior of the thermal Drude weight we can then conjecture that it is
given by Eqs.~(\ref{eq:mf7}) and (\ref{eq:mf9}) for $h=h_{c2}$, independently of $\Delta$. 
We will come back to this issue in Sec.~\ref{sec:4} where we discuss 
the results from exact diagonalization for $\Delta>0$. 
The case of $\Delta=-1$ and $h=0$, however,  
seems to be  an exception as we have found 
indications for $K_{\mathrm{th}}(h=0)\propto T$ at low temperatures before\cite{hm1}.
Here, the existence of many   low-lying excitations might complicate  the situation.
\\\indent
 In the ferromagnetic state and for low temperatures,
the  parameter   $\alpha$ from Eq.~(\ref{eq:mf2}) is  exponentially suppressed and the average local 
magnetization is $m=1/2$. Thus, to leading order  in $T$ the low-temperature 
dependence of $K_{\mathrm{th}}(h,T)$ is 
independent of $\Delta$, similar to the case of $h=h_{c2}$, and the thermal Drude weight is exponentially
suppressed
$
K_{\mathrm{th}}(h,T)\propto T^{3/2} e^{-G/T}
$ with $G=h-h_{c2}$.\\\indent
In the gapless state our mean-field theory results confirm that $K_{\mathrm{th}}(h,T)= V(h,\Delta)\, T$ for $\Delta>0$
and low temperatures. 
However,  the mean-field prefactor $V(h,\Delta)$ will be renormalized if
 interactions are fully accounted for; see Eq.~(\ref{eq:xx}).\\
\indent
In summary, we have obtained the leading low-temperature contributions to
$K_{\mathrm{th}}$ in the regimes (i) and (ii) of Table~\ref{tab:1} using
mean-field theory and conformal field theory. Mean-field theory provides  a reasonable
quantitative description of the transport
coefficients for small $\Delta$ and $h$ as well as for $h\geq h_{{\mathrm{c2}}}$.


\section{Exact diagonalization}\label{sec:4}
\begin{figure}[t]
\centerline{\epsfig{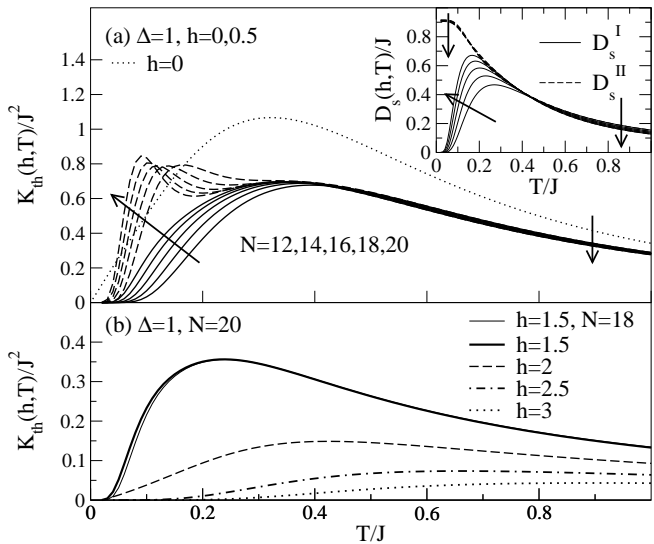}}
\caption{
Thermal Drude weight $K_{\mathrm{th}}(h,T)$ of the Heisenberg chain ($\Delta=1$).
Panel (a): $K_{\mathrm{th}}(h,T)$ computed from Eq.~(\ref{eq:k5a}) using
$D_{\mathrm{s}}=D_{\mathrm{s}}^{I}$ (solid lines) and 
$D_{\mathrm{s}}=D_{\mathrm{s}}^{II}$ (dashed lines), both for $h=0.5$ and
$N=12,14,16,18,20$. Arrows indicate increasing system size. 
The dotted line is the result for $h=0$ and $N\to \infty$ from Ref.~\onlinecite{kluemper02}.
Inset: comparison of  $D_{\mathrm{s}}^{I}(h,T)$ [solid lines] and 
$D_{\mathrm{s}}^{II}(h,T)$ [dashed lines]; $h=0.5$, $\Delta=1$. 
Panel (b): $K_{\mathrm{th}}(h,T)$ for $h=1.5,2,2.5,3$ and $N=20$ [thick lines;
$D_{\mathrm{s}}(h,T)=D_{\mathrm{s}}^{I}(h,T)$].
The curve for $N=18,h=1.5$ is included (thin solid line). }\label{fig:2}
\end{figure}

In this section, we first present numerical results for the thermal Drude weight of the Heisenberg chain ($\Delta=1$).
Second,  the  field dependence of $K_{\mathrm{th}}(h,T)$ for intermediate temperatures $T$ is analyzed.
Next, $K_{\mathrm{th}}(h,T)$ for $h=h_{c2}$ is discussed for different choices of the anisotropy $\Delta\geq 0$
and finally,
we make some remarks on the lower bound for the  spin Drude weight $D_{11}=D_{\mathrm{s}}$ given in Eq.~(\ref{eq:21}). 
While $D_{\mathrm{s}}(h,T)$ still eludes an exact analytical treatment for arbitrary temperatures,
 analytically exact results for $D_{\mathrm{th}}(h,T)$ and $D_{\mathrm{th,s}}(h,T)$ of the Heisenberg chain have 
  very recently been reported
in Ref.~\onlinecite{sakai04}.
\\
\indent
Let us first address a technical issue, namely the appropriate choice for
$D_{\mathrm{s}}(h,T)$   in Eq.~(\ref{eq:k5a}). 
For the case of zero magnetic field, we know 
from our previous study 
Ref.~\onlinecite{hm2} that 
$D_{\mathrm{s}}^{I}(h,T)$
and $D_{\mathrm{s}}^{II}(h,T)$ exhibit a different finite-size behavior at $h=0$.
This is similar to the situation at finite fields. The  inset of Fig.~\ref{fig:2}(a)
shows both  $D_{\mathrm{s}}^{I}(h,T)$ and $D_{\mathrm{s}}^{II}(h,T)$ for $\Delta=1$ and $h=0.5$,
and we see that first,  $D_{\mathrm{s}}^{II}(h,T)$ is well converged at low temperatures; 
and second, a large difference between $D_{\mathrm{s}}^I(h,T)$ and 
$D_{\mathrm{s}}^{II}(h,T)$ is visible at low temperatures. The thermal Drude weight $K_{\mathrm{th}}(h,T)$,
resulting from either inserting $D_{\mathrm{s}}^{I}(h,T)$ or $D_{\mathrm{s}}^{II}(h,T)$ in Eq.~(\ref{eq:k5a}),
is shown in Fig.~\ref{fig:2}(a). We have decided to use $D_{\mathrm{s}}^{I}$ in the  numerical study for consistency reasons, 
since then, all Drude weights entering in Eq.~(\ref{eq:k5a}) have a similar finite-size dependence at low temperatures,
characterized by the exponential suppression at low temperatures due to the finite-size gap. On the contrary, using
 $D_{\mathrm{s}}^{II}(h,T)$ leads to an artificial  double peak structure in $K_{\mathrm{th}}(h,T)$; 
 seen in Fig.~\ref{fig:2}(a).

\begin{figure}[t]
\centerline{\epsfig{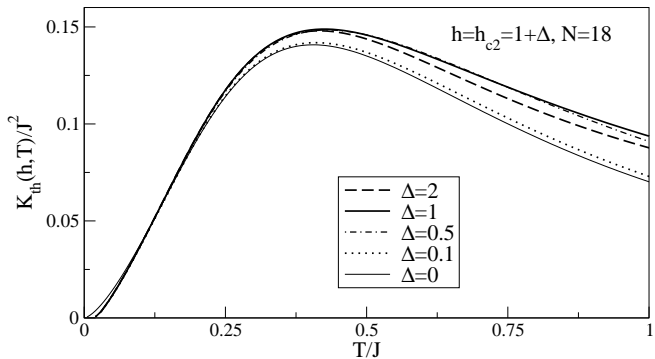}}
\caption{
Thermal Drude weight $K_{\mathrm{th}}(h,T)$ at the critical field $h_{c2}=1+\Delta$ 
for $\Delta=0,0.1, 0.5, 1,2$. 
For $\Delta\not= 0$, we show numerical results for $N=18$ sites, while the curve for 
the free fermion case  (thin solid line) is valid in the thermodynamic limit.
}\label{fig:4}
\end{figure}

\begin{figure}[t]
\centerline{\epsfig{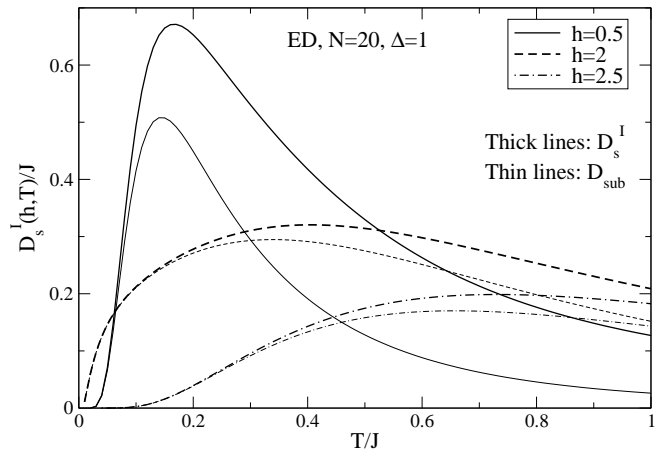}}
\caption{Comparison of the spin Drude weight $D_{\mathrm{s}}^I(h,T)$ (thick lines) and its lower bound
$D_{\mathrm{sub}}$
(thin lines; see text for further details). The figure shows results for $\Delta=1$, $N=20$, and $h=0.5,2,2.5$.
}\label{fig:5} 
\end{figure}

 We have checked that a similar scenario arises  for  $\Delta=0$ for finite systems. 
 However, for this case the Drude weight can be computed exactly in the thermodynamic limit and
 we find that one of the two maxima disappears.
 Thus we  expect an
 analogous behavior for $\Delta>0$, supporting the choice of $D_{\mathrm{s}}^I$ instead of $D_{\mathrm{s}}^{II}$.
\\\indent
Further numerical results for $K_{\mathrm{th}}(h,T)$ of the Heisenberg chain are 
provided in Fig.~\ref{fig:2}(b)
for $h\geq 1.5$. 
The main features of the thermal Drude weight can be summarized as follows: 
(i) for $0<h<h_{c2}$, finite-size effects   are  small for $T/J\gtrsim 0.4$ \lbrack see 
Fig.~\ref{fig:2}(a)\rbrack; 
(ii) for $h\geq h_{c2}$, finite-size effects are negligible;
(iii) the position of the maximum depends on  the magnetic field;
(iv)  $K_{\mathrm{th}}(h,T)$ is strongly suppressed as the magnetic field is
increased.
\\
\indent
As both $D_{\mathrm{th}}(h,T)$ and $D_{\mathrm{th,s}}(h,T)$ converge rapidly to the thermodynamic limit at high temperatures, 
the small finite-size effects observed for $T/J\gtrsim 0.4$ [see Fig.~\ref{fig:2}(a)] are due to $D_{\mathrm{s}}(h,T)$ 
[see the inset of Fig.~\ref{fig:2}(a)]. At low $T$, $K_{\mathrm{th}}(h,T)$ increases with system size $N$ while it decreases
with growing $N$ 
at high temperatures.
The vanishing of pronounced finite-size effects upon approaching the line $h=h_{c2}$ from below 
can be ascribed to  the  fact that a description in terms of free fermions
with parameters independent of $\Delta$ is valid here,
as was already evidenced in the previous section.
For the ferromagnetic state ($h>h_{c2}$), the curves shown in Fig.~\ref{fig:2}(b) for $N=20$ are  
indistinguishable
from the corresponding ones for $N=18$ (not included in the figure) within the line width. \\
\begin{figure}[t]
\centerline{\epsfig{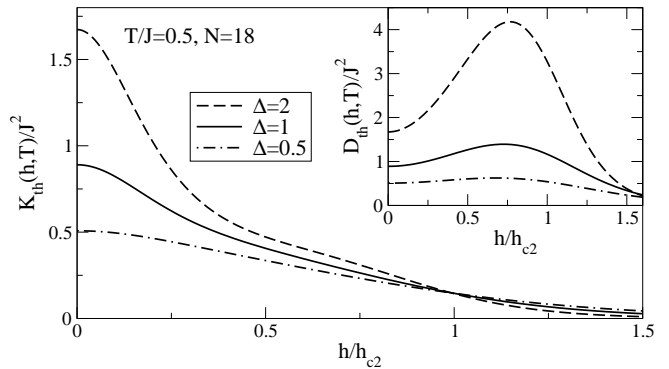}}
\caption{
Main panel: field dependence of the thermal Drude weight $K_{\mathrm{th}}(h,T)$ for $\Delta=0.5,1,2$ and $T/J=0.5$
(ED for $N=18$ sites). Inset: field dependence of $D_{\mathrm{th}}(h,T)$ for the same parameter sets as in the main
panel.}\label{fig:3}
\end{figure}
\indent
Regarding the position of the maximum, there is evidence that it is first
shifted to higher temperatures when the field is switched on
as compared to the case of $h=0$; see Fig.~\ref{fig:2}(a).
 A precise determination of its 
position in the intermediate 
gapless phase is somewhat complicated as  typically, the numerical data converge 
well down to roughly only the peak temperature. Still, there are indications that at strong fields $h\sim 1$,
the maximum tends to be located at lower temperatures than for
$h=0$. This can be seen, for instance,  in the case of $h=1.5$
in Fig.~\ref{fig:2}(b).
In the polarized state, $K_{\mathrm{th}}(h,T)$ definitely peaks at larger temperatures than
at vanishing field due to its exponential suppression at low temperatures.\\
\indent 
The decrease of $K_{\mathrm{th}}(h,T)$ as a function of increasing magnetic
field as mentioned 
in the preceding discussion of the Heisenberg chain is also observed 
for other choices for the anisotropy $\Delta$. This is demonstrated for $\Delta=0.5,1,2$ at $T/J=0.5$ in 
the main panel of Fig.~\ref{fig:3}, where $K_{\mathrm{th}}(h,T)$ is shown as a function of the magnetic field $h$
and plotted versus $h/h_{c2}$. 
In contrast to $K_{\mathrm{th}}(h,T)$, $D_{\mathrm{th}}(h,T)$ grows   with increasing magnetic field 
at intermediate temperatures. 
For illustration,
$D_{\mathrm{th}}(h,T)$ is plotted versus $h/h_{c2}$ at $T/J=0.5$ in the inset of Fig.~\ref{fig:3} for the same choice of parameters
as in the main panel.
It exhibits a maximum at large fields, which increases 
and its position approaches $h=h_{c2}$ when
the temperature is lowered. Thus, indications of the transition to the ferromagnetic phase are 
visible in $D_{\mathrm{th}}(h,T)$, but not present in $K_{\mathrm{th}}(h,T)$. Note, however, that 
all three curves in the main panel of Fig.~\ref{fig:3} almost pass  through the same point for 
$h\approx h_{c2}$.
 For $T/J\lesssim 1$, $D_{\mathrm{th}}(h,T)$  is  enhanced by the magnetic field and we can therefore conclude that the
decrease of $K_{\mathrm{th}}(h,T)$ as a function of magnetic field is due to a cancellation of $D_{\mathrm{th}}(h,T)$ 
and the magnetothermal correction in Eq.~(\ref{eq:k5a}).\\
\indent
Along the critical line $h=h_{c2}=1+\Delta$, further evidence for 
universal low-temperature behavior can be found by ED. 
This can be seen in Fig.~\ref{fig:4}, where we present $K_{\mathrm{th}}(h,T)$ for
$\Delta=0.1,0.5,1,2$ and $N=18$. The curve for $\Delta=0$ is also included in the figure; this one,
however,  is exact in the thermodynamic limit. Below $T/J\approx 0.25$, the curves lie on top of each
other. Small deviations at lowest temperatures visible in the plot can be ascribed to the presence of finite-size gaps.
This supports our conclusion from Sec.~\ref{sec:3} that Eqs.~(\ref{eq:mf7}) and (\ref{eq:mf9}) hold for arbitrary $\Delta\geq 0$ and further numerical 
data (not included in the figure) show that it is also correct for $-1<\Delta<0$. \\
\indent 
Finally, let us turn to the inequality Eq.~(\ref{eq:21}) for the spin Drude
weight $D_{11}(h,T)=D_{\mathrm{s}}(h,T)$ introduced in Sec.~\ref{sec:2}.
Here, we want to  discuss to which extent the inequality Eq.~(\ref{eq:21}) is exhausted 
by $j_{\mathrm{th}}$ at finite magnetic fields and finite temperatures. An analogous analysis  
in the limit of  $\beta=0$ can be found in 
Ref.~\onlinecite{zotos97}. 
  To this end we compare  $D_{\mathrm{s}}^{I}(h,T)$ and 
 \begin{equation}
 D_{\mathrm{sub}}(h,T):=\frac{\pi}{T\,N} \frac{\langle j_{\mathrm{s}}j_{\mathrm{th}}\rangle^2}{\langle  j_{\mathrm{th}}^2\rangle}
 =\frac{1}{T}\frac{D_{\mathrm{th,s}}^2(h,T)}{D_{\mathrm{th}}(h,T)}
\label{eq:22}
\end{equation}
in Fig.~\ref{fig:5}. Note that first, the relation $D_{\mathrm{s}}(h,T)\geq D_{\mathrm{sub}}(h,T)$ is equivalent to the positivity
of the thermal Drude weight $K_{\mathrm{th}}(h,T)\geq 0$. Second,  $D_{\mathrm{s}}(h,T)\approx D_{\mathrm{sub}}(h,T)$ implies a very small
thermal Drude weight and thus, the comparison provided in Fig.~\ref{fig:5} also reveals the relative size
of the two contributions to $K_{\mathrm{th}}(h,T)$ in Eq.~(\ref{eq:k5a}), namely $D_{\mathrm{th}}(h,T)$ and the magnetothermal correction
$D_{\mathrm{th,s}}^2(h,T)/\lbrack T\, D_{\mathrm{s}}^I(h,T)\rbrack $. In Fig.~\ref{fig:5}, results are shown for $\Delta=1$,
$N=20$ sites, and
$h=0.5,2,2.5$. For the sake of clarity, data for smaller system sizes are not included in the figure. Differences between
the curves for $N=18$ and $N=20$ are anyway only pronounced  for temperatures $T/J\lesssim 0.1$ 
and become smaller
as the magnetic field $h$ increases.\\
\indent
Figure \ref{fig:5} allows for three major observations: 
(i)   $D_{\mathrm{s}}^I(h,T)\approx D_{\mathrm{sub}}(h,T)$ at low temperatures and for all cases shown in the figure;
(ii)  $D_{\mathrm{sub}}(h,T)$ approximates $D_{\mathrm{s}}^I(h,T)$  the better the larger the magnetic field is;
(iii) significant deviations are present for high temperatures implying that for a quantitative description of 
$D_{\mathrm{s}}(h,T)$ using Eq.~(\ref{eq:21}),
more conserved quantities need to be considered in Eq.~(\ref{eq:21}).\\
\indent
Our comparison provides, at least for finite system sizes, a quantitative measure
of the temperature range where $D_{\mathrm{s}}^I\approx D_{\mathrm{sub}} $.
Point (i) indicates that analytical approaches can
make use of $D_{\mathrm{sub}}(h,T)$ 
for a quantitative description of $D_{\mathrm{s}}(h,T)$  at low temperatures as it has been done
by Fujimoto and Kawakami within a continuum theory in Ref.~\onlinecite{fujimoto03}. 
The quantities that appear on the right hand side of Eq.~(\ref{eq:22})
are less involved than Eqs.~(\ref{eq:k3}) and (\ref{eq:k8}) for $D_{\mathrm{s}}(h,T)$, as the
former are static correlators.
Furthermore, for finite magnetic fields, we suggest to compute $D_{\mathrm{s}}(h,T)$ analytically from Eq.~(\ref{eq:21}),
taking into account some of the  conserved quantities $Q_m$, which are in principle  known (see, e.g., Ref.~\onlinecite{grabowski96}). 
 Such a procedure is applicable to $h\not= 0$ and might circumvent the 
ambiguities in the results encountered in recent  computations of
$D_{\mathrm{s}}(h=0)$ (Refs.~\onlinecite{zotos99,long03,fujimoto03,akluemper}). The latter 
have used Eq.~(\ref{eq:k8}) directly or Kohn's formula\cite{kohn64,castella95},
equivalently. Regarding the relative size of $D_{\mathrm{th}}(h,T)$ and the magnetothermal correction, 
we see that the latter becomes more relevant
the larger the magnetic field is which leads to the strong suppression of
$K_{\mathrm{th}}(h,T)$. This is consistent with results of the 
previous sections of this paper.


\section{Conclusions}\label{sec:5}
In this paper we have studied the thermal Drude weight of the $XXZ$ model with exchange anisotropy $\Delta\geq 0$
in finite magnetic fields using mean-field theory and exact diagonalization. Magnetothermal effects have been taken into 
account and the condition of zero magnetization current flow has been applied. Let us now summarize the main findings and 
relate them to experiments.\\
\indent
We have discussed  the low-temperature limit of the thermal Drude weight $K_{\mathrm{th}}(h,T)$ 
and we have given arguments that it changes from an algebraic
behavior for $0 \leq\Delta \leq 1, h\leq h_{c2}$  to an exponentially activated behavior in the polarized state
for $h>h_{c2}$. In addition, the leading term at low temperatures along the critical 
line $h=h_{c2}, \Delta > -1$, is universally given by $K_{\mathrm{th}}(h,T)=
A\, T^{3/2}$, where the prefactor $A$, given in Eqs.~(\ref{eq:mf7}) and (\ref{eq:mf9}), is independent of $\Delta$. 
In the gapless phase, the leading contribution to  $K_{\mathrm{th}}(h,T)$ is linear in the temperature with a field- and anisotropy
dependent prefactor.
In consequence,
the thermal Drude weight $K_{\mathrm{th}}(h,T)$ can be expected to be  proportional to the specific heat in the gapless state
in the low-temperature limit, where the velocity of elementary excitations is constant.\\ \indent
Further, the Drude weight is suppressed by the magnetic field, which can be ascribed to the increase of the 
magnetothermal correction relative to the pure thermal Drude weight $D_{\mathrm{th}}(h,T)$. 
As a third result, the position of the maximum of $K_{\mathrm{th}}(h,T)$ depends
non-monotonically on
the magnetic field. While in the present paper, we have focused on the thermal Drude weight $K_{\mathrm{th}}$ under the 
condition of zero spin-current flow, our analysis of the full transport matrix Eq.~(\ref{eq:k0}) 
can  easily be extended to a variety of other transport situations, which would be
characterized by different combinations of the Drude weights.\\
\indent
Turning now to experiments, we emphasize that for a description of realistic materials external scattering has to be accounted
for. In a simple picture, one may expect the Drude peak to be broadened by  external 
scattering mechanisms. The behavior in magnetic fields that one may observe in experiments will
likely depend both on external scattering rates as well as on the thermal Drude weight. 
Nevertheless, one may speculate that qualitative trends of the field dependence of the thermal Drude weight 
are  reflected in thermal transport experiments.
\\\indent
In recent experiments on quasi one-dimensional magnetic materials\cite{solo00,solo01,kudo01,hess01,solo03}, the thermal conductivity has often been found 
to be insensitive to the application of an external magnetic
field. This is, however, explained by the large absolute value of the exchange coupling in these materials, being typically of the
order of magnitude of $1000$ K. It would therefore be desirable to perform measurements on materials with a moderately
small exchange coupling to check our results. Still, it might be very difficult to reach the saturation field $h_{c2}$
in realistic materials, but at least the qualitative features like the suppression of the thermal conductivity or a shift of the 
maximum could be verified. In particular, experiments in relatively large magnetic fields may provide an indirect probe of
spin currents in one-dimensional quantum magnets, which have so far not yet been observed directly in experiments. Conclusions about the low-temperature limit
would require reliable methods to separate the    magnetic contribution from the phonon part, 
which is a challenging task in the interpretation of
experiments. Nevertheless, we believe that further experiments on the field dependence   of the thermal conductivity
could hint at the nature and mechanisms of magnetic transport properties.

\indent {\bf Acknowledgments - }
This work was supported by the DFG, Schwer\-punkt\-programm 1073.
 It is a pleasure to thank 
B.~B\"uchner, D.C.~Cabra,  and C.~Hess
for fruitful discussions. We are indebted to C.~Hess for a careful reading of
the manuscript and valuable suggestions.
We acknowledge support by the Rechenzentrum of the TU Braunschweig
where
parts of the numerical computations have been performed on a COMPAQ ES45.


\end{document}